\documentclass[aps,prd,reprint,superscriptaddress]{revtex4-2}
\usepackage[english]{babel}
\usepackage[utf8]{inputenc}
\usepackage[colorinlistoftodos, color=green!40, prependcaption]{todonotes}
\usepackage{orcidlink}
\usepackage{amsthm}
\usepackage{mathtools}
\usepackage{physics}
\usepackage{xcolor}
\usepackage{graphicx}
\usepackage{adjustbox}
\usepackage{placeins}
\usepackage[T1]{fontenc}
\usepackage{lipsum}
\usepackage{csquotes}
\usepackage{hyperref}
\usepackage{lipsum}

\begin{document}
\title{511 keV Gamma Ray Echo from Particle Decays in Supernovae}
\author{Garv Chauhan\,\orcidlink{0000-0002-8129-8034}}
   \affiliation{Department of Physics, Arizona State University, 450 E. Tyler Mall, Tempe, AZ 85287-1504 USA}
\author{Cecilia Lunardini\,\orcidlink{0000-0002-9253-1663}}
    \affiliation{Department of Physics, Arizona State University, 450 E. Tyler Mall, Tempe, AZ 85287-1504 USA}

\begin{abstract}
The formation of a hot and dense core in a core-collapse supernova (SN) can produce massive Beyond Standard Model (BSM) particles. These particles can decay in the stellar envelope, generating positrons either directly or through secondary processes involving neutrinos or photons. We show for the first time that such positrons regardless of their production channel, can thermalize and annihilate at rest with ambient electrons in the outer SN envelope, producing a characteristic \emph{echo} of 511 keV gamma rays. For axion-like particles (ALPs), we derive bounds on the ALP–photon coupling ($G_{a\gamma}$) using Pioneer Venus Orbiter observations of SN 1987A. We also evaluate the sensitivity of upcoming MeV gap gamma-ray telescopes in the 511 keV range, such as COSI and AMEGO, for future Galactic SNe, which can improve existing constraints or enable ALP discovery. The echo signal is a generic prediction for any particle species that efficiently produces positrons near the stellar surface.
\end{abstract}

\maketitle

\textit{Introduction.}$-$ Core-collapse supernovae (CCSNe) represents powerful laboratories for probing the physics Beyond the Standard Model (BSM)~\cite{Raffelt:1996wa}. Indeed, a significant fraction of the energy released in the core collapse could be emitted in the form of BSM particles -- like axion-like particles (ALPs), sterile neutrinos, majorons, and more -- in addition to the dominant cooling channel of neutrino emission. The existence and properties of these exotic particles could manifest themselves as missing energy in the dominant channel, or by direct observation of the exotica themselves or their Standard Model byproducts. Observations of the last nearby supernova, SN1987A \cite{Kamiokande-II:1987idp,Bionta:1987qt}, remains a unique benchmark of these methods (see, e.g.  \cite{Falk:1978kf,Choi:1987sd,Kolb:1988pe,Chupp:1989kx,Raffelt:1990yz,Oberauer:1993yr,Jaffe:1995sw}). For a future supernova in or near our galaxy, improved observation capabilities in multimessenger astronomy -- including neutrinos, photons, cosmic rays and gravitational waves -- carry the promise of a greater constraining and discovery power of BSM particles.  

In this multimessenger landscape, gamma ray telescopes play an important role. They can be used to search for anomalous photon flashes that could occur due to exotic particles decays during the ``dark'' period between core collapse and shock breakout, when the stellar envelope is not yet disturbed by the aftermath of the collapse. For example, BSM constraints were obtained from the non-observation of high-energy gamma-rays from SN1987A in the Gamma Ray Spectrometer (GRS) aboard the Solar Maximum Mission (SMM)~\cite{Kolb:1988pe,Chupp:1989kx,Oberauer:1993yr} and Gamma Ray Burst detector aboard the
Pioneer Venus Orbiter~\cite{Jaffe:1995sw}. 
%
Until recent times, in the region between the hard X-ray and high-energy gamma-ray regimes, at $E\sim 0.1 - 100 $ MeV, surveys have suffered from poor resolution and sensitivity. This energy range termed the ``\text{MeV gap}'' is crucial to study important astrophysical processes such as nuclear line emission, positron annhilation and gamma-ray bursts.  A number of upcoming and next-generation telescopes aim at improving  line sensitivity, lower instrumental backgrounds and  achieve large effective areas.
Examples are COSI \cite{Tomsick:2021wed}, which is under active development at NASA  (launch planned for 2027), and the longer term projects e-ASTROGAM \cite{e-ASTROGAM:2017pxr} and AMEGO \cite{Kierans:2020otl}, which will reach impressive effective areas of $100$-$1000$ cm$^2$.

\vspace{2pt}
\begin{figure}[t!]
    \centering
    \includegraphics[width=\linewidth]{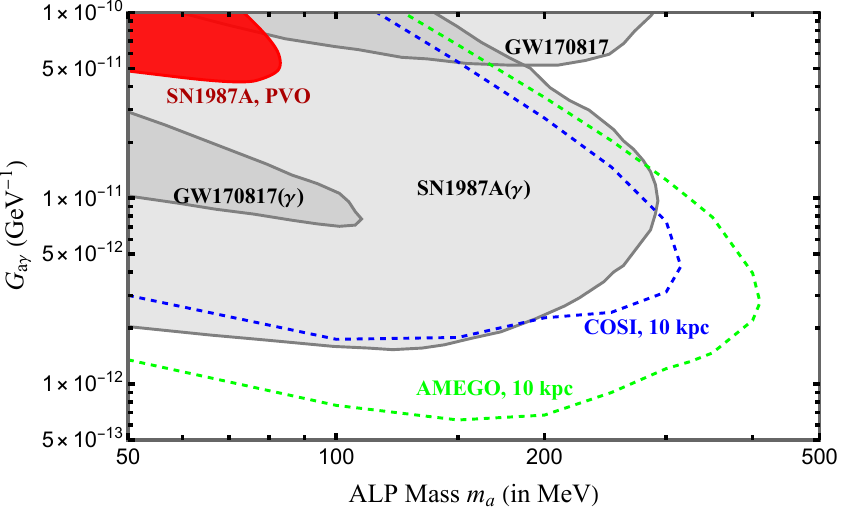}
    \caption{Constraints in the $G_{a\gamma}$-$m_a$ space from PVO observations during SN1987A (red-shaded region) and future sensitivities of COSI (blue-dashed region) and AMEGO (green-dashed region) for a SN at 10 kpc. The limits derived in our work correspond to a $3\sigma$ confidence interval. We also show existing bounds, from Refs.~\cite{Jaeckel:2017tud,Hoof:2022xbe,Muller:2023vjm,Dev:2023hax,Diamond:2023cto} (details in the text).}
    \label{fig:MainConstraint}
\end{figure}
In this \textit{Letter}, we propose a novel BSM signature in gamma rays in the MeV gap. 
We show that any in-medium BSM decays that produces positrons -- either directly or through intermediate processes involving neutrinos or photons -- will lead to the production of a unique 511 keV line signal from positron annihilation at rest. Some of these photons will reach Earth with some time delay relative to the $\sim 10$ s long neutrino burst, and -- to highlight  its association with new particles emitted from the SN core, we term their flux at Earth as the BSM \textit{Gamma Ray Echo} of a core collapse supernova. As we will show, the luminosity of this echo can greatly surpass the analogous signal due to purely Standard Model processes, mainly positrons produced by electron antineutrino interactions in the hydrogen envelope of the star i.e. Inverse Beta Decay (IBD). For the latter, we refer to \cite{BisnovatyiKogan:1975,Ryazhskaya:1999} for early studies, and to  \cite{Lu:2007wp,Lunardini:2023ilg} for more modern treatments, from which useful results will be used here.

We also stress that our echo signal differs from the diffuse 511 keV signal studied earlier~\cite{DeRocco:2019njg,Calore:2021klc,Calore:2021lih}. In these studies, the signal arises from the particle decays directly to positrons and needs to occur outside the stellar envelope. The positron annihilation into 511 keV gamma-rays occurs in the interstellar medium over long-time scales of about $10^{5}$-$10^{6}$ years. Our work can be extended to study such positron insertion into the interstellar medium from species decaying to high-energy photons and neutrinos.

\begin{figure}[t]
    \centering
    \includegraphics[width=\linewidth]{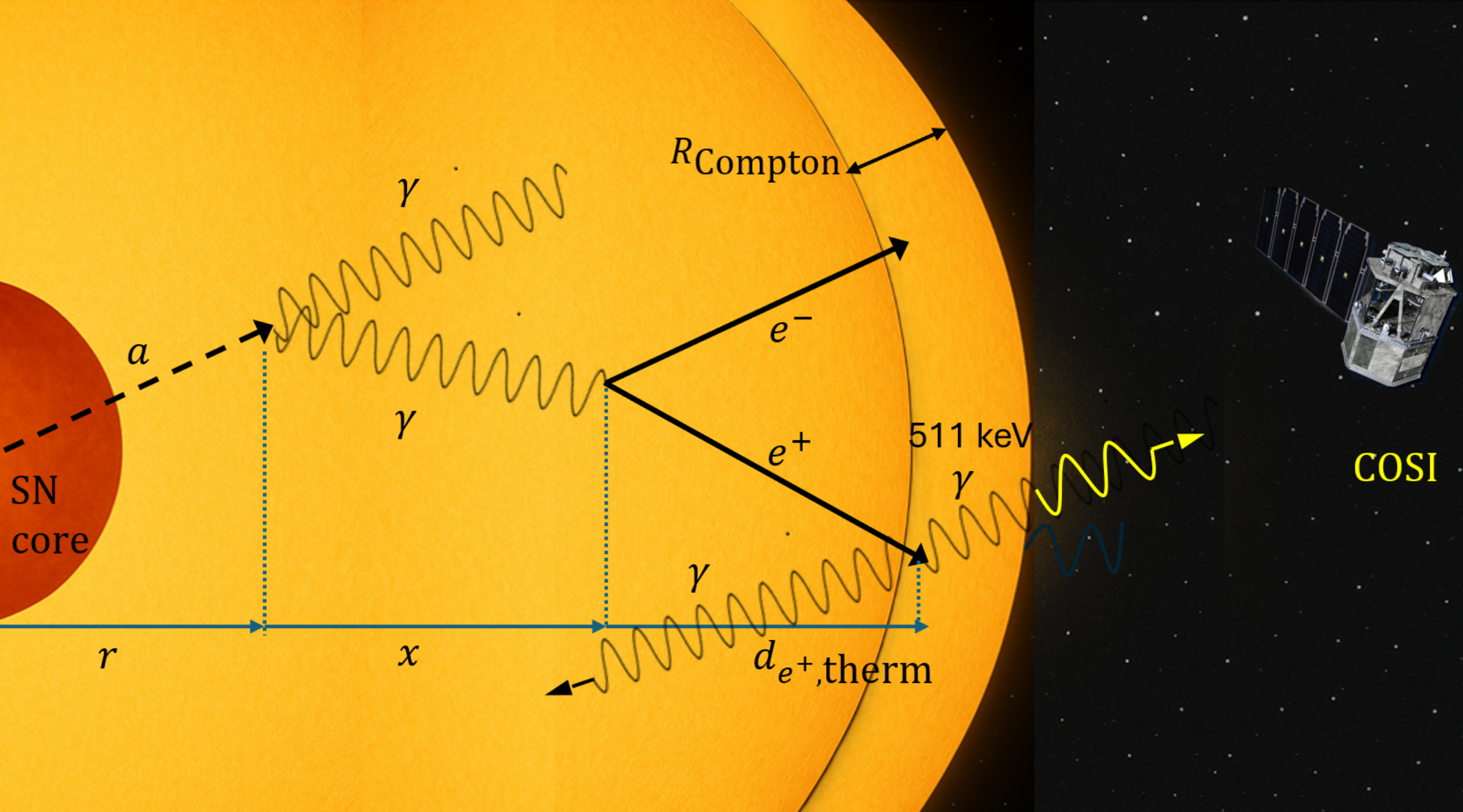}
    \caption{A schematic representation of the formation of a 511 keV signal from the ALP decay chain, Eq. (\ref{eq:gammadecay}). See also Eq. (\ref{eq:diffNgamma}).}
    \label{fig:ALPcartoon}
\end{figure}

\textit{Decay Modes.}$-$ 
The 511 keV line emission crucially depends on the annihilation of the thermalized positrons. This required positron flux can be generated in three salient ways,
\begin{equation}
     \text{X}\xrightarrow{\text{decay}}e^+ +\text{anything}
\end{equation}
\begin{equation}
    \text{X}\xrightarrow{\text{ decay }}\gamma+\text{anything};\quad\gamma\xrightarrow{\text{pair prod.}}e^-e^+
    \label{eq:gammadecay}
\end{equation}
\begin{equation}
    \text{X}\xrightarrow{\text{ decay }}\bar{\nu}_e+\text{anything};\quad\bar{\nu}_e+p\xrightarrow{\text{IBD}}n+e^+
\end{equation}
where $\text{X}$ denotes any generic BSM species. In this work, we will only study the generation of 511 keV signal from BSM particle decays into high-energy photons. The mechanism and calculation for particles decaying  into neutrinos remains qualitatively similar to the SM case (e.g., \cite{Lunardini:2023ilg}), with few differences arising in the light curve properties from the temporal delays due to the finite mass of the BSM species. The calculation for the particles decaying directly into a electron-positron pair will be similar to our discussion for the high-energy photon case.  

To illustrate the case of BSM species decaying into a pair of high-energy photons, we focus on the ALP coupling to the electromagnetic field with the relevant Lagrangian 
\begin{equation}
    \mathcal{L}_{a\gamma} \supset -\frac{1}{4}G_{a\gamma}F_{\mu\nu}\Tilde{F}^{\mu\nu}\, a~,
    \label{eq:lagrphoton}
\end{equation}
where $G_{a\gamma}$ is the ALP-photon coupling constant, $F_{\mu\nu}$ is the electromagnetic field strength tensor, $a$ is the ALP field and with decay length given by
\begin{equation}
\lambda_{a\rightarrow\gamma\gamma}=\frac{64\pi}{G_{a\gamma}^2 m_a^4}\sqrt{E_a^2-m_a^2}~.
\label{eq:lambdaphoton}
\end{equation}
where $m_a$ is ALP mass and with ALP energy $E_a$. In the ALP's rest frame, the decay into two photons is isotropic, implying that the energy distribution of the daughter particles in the lab frame is a box function i.e. $E_{\gamma}\in [\gamma_a(1-\beta_a)m_a/2,\gamma_a(1+\beta_a)m_a/2]$, where $\gamma_a=E_a/m_a$ is the Lorentz boost factor. For $E_a\gg m_a$, the differential angular distribution of the photons is forward peaked (see supplemental material).

\textit{Reference SN model \& ALP Production.}$-$ 
As a reference for our calculations, we use the results of the 1-dimensional 18.8 $M_\odot$ SN simulation (SFHo-18.8) by the Garching group performed with the \texttt{PROMETHEUS VERTEX} code~\cite{SNprofile,Mirizzi:2015eza,Bollig:2020xdr}.  The SN core radius is set to $R_{\text{core}}\sim 20$ km for BSM species production. For consistency, to model the ALP decay chain we adopt the density profile of the progenitor star used in the SFHo-18.8 simulation, where the envelope radius is $R_{\text{env}} \sim 6 \times 10^{12}$ cm.  We take this profile to be time-independent  because, for ALPs, the major contribution from decays originate at most $\sim 10^{11}$ cm deep inside the stellar surface (see supplemental material), and this region remains undisturbed until shock breakout, which we estimate to be about $\sim 6-9$ hours after collapse. This time scale is longer than the scale of ALP emission and subsequent decay chain. We assume the composition of the envelope to be hydrogen dominated with average $Y_p=0.7$ and $Y_e=0.85$. 

Let us discuss ALP production near the collapsed supernova core. For $G_{a\gamma}$ case (eq. (\ref{eq:lagrphoton})), the ALP production occurs primarily through two processes, Primakoff upscattering and photon coalescence process. In Primakoff upscattering, production occurs through a photon exchange with the background charged medium ($\gamma+Ze\rightarrow a+ Ze$, where $Z$ denotes the species charge). The rate depends on the effective density of charged particles, with protons giving the dominant contribution. Ignoring the photon plasma mass, the number rate per unit volume is~\cite{Raffelt:1985nk,Dicus:1979ch,DiLella:2000dn,Lucente:2020whw}
\begin{equation}
    \left.\frac{dn_a(r,E_a^*,t^*)}{dE_a^*}\right|_{\text{P}}= \hat{n}\frac{\alpha G_{a\gamma}^2}{2\pi^2}\frac{{E_a^*}^2}{e^{E_a^*/T}-1}f_\text{P}
    \label{eq:sub1}
\end{equation}
where $E_a^*$ and $t^*$ are the ALP energy and elapsed time in the PNS core frame respectively, $T$ is the temperature of the SN core, $\hat{n}$ is the effective charged-particle density and $f_\text{P}$ is a function dependent on the screening scale, plasma frequency and the ALP energy (see supplemental material).

The production through coalescence of two-photons ($\gamma+\gamma\rightarrow a$) is enhanced for heavier ALP masses because the production rate scales like $m_a^4$ as can be seen clearly from the emission rate~\cite{Lucente:2020whw,Caputo:2022mah,Ferreira:2022xlw}  
\begin{equation}
    \left.\frac{dn_a(r,E_a^*,t^*)}{dE_a^*}\right|_{\text{C}}= \frac{G_{a\gamma}^2\,m_a^4}{128\pi^3}\sqrt{{E_a^*}^2-m_a^2}\,e^{-E_a^*/T}f_\text{B}
    \label{eq:sub2}
\end{equation}
where $f_\text{B}$ is the averaged stimulation factor for the two photons dependent on the SN core temperature, ALP mass and energy (see supplemental material).

We computed the number luminosity of ALPs produced inside the SN core upto radial coordinate $r=R_{\text{core}}$ (with $r=0$ being the center of the star, which is assumed to be spherically symmetric), in per unit energy $d^2N_a(r,E_a,t)/dt\,dE_a$~\cite{Ferreira:2022xlw}
\begin{align}
   \frac{d^2N_a(r,E_a,t)}{dt\,dE_a} &= \int_0^{R_\text{core}} dr\,4\pi r^2\,\frac{dn_a(r,\eta_{\text{lapse}}^{-1}E_a,\eta_{\text{lapse}}t)}{dE_a^*} 
   \label{eq:main1}
\end{align}
where we include the effects of gravitational red-shifting by including the lapse factor $\eta_{\text{lapse}}(r)$. Therefore, ALP energy and elapsed time $t$ for an observer far away from the SN core is $E_a = \eta_{\text{lapse}} E_a^*$ and $t = \eta_{\text{lapse}}^{-1} t^*$ respectively.

\textit{Pair-Production.}$-$In the ALP-photon case, the energetic $\gamma$-rays interact with the plasma protons and electrons in the stellar envelope, which enables pair-production ($\gamma \rightarrow e^+ e^-$). In the absence of matter, a dense photon field can exist in certain regions of the parameter space, potentially leading to the formation of a thermal plasma consisting of $\gamma,e^\pm$ through $\gamma\gamma\rightarrow e^+e^-$ i.e. fireball scenario~\cite{Diamond:2023scc}. However, pair production on matter is generally stronger, as is the case in our scenario. The photon mean free path (m.f.p.), $\lambda_{\text{pair}}$ in matter, depends on the Bethe-Heitler cross-section ($\sigma_{\mathrm{pair}}$) and the number density of the spectator charged species. For hydrogen (number density $n_{\text H}$), it is $\lambda^{-1}_{\text{pair}}(E_\gamma; x) = n_{\text H}\,\sigma_{\mathrm{pair}}(E_\gamma)$.  The positron energy distribution, $g_{{e^+}}(E_{e^+},E_\gamma)$, can be estimated from the differential cross section~\cite{Bethe:1934za,Tsai:1973py,ParticleDataGroup:2024cfk} (see supplemental material). In our calculation, we are restricted to the region near the stellar surface which is transparent to 511 keV photons and its immediate surroundings (see below for details).  There, most of the hydrogen near the stellar surface is ionized ~\cite{Lu:2007wp}, therefore we simply use the expression of $\sigma_{\mathrm{pair}}$ for the case of unscreened nuclear field \cite{Bethe:1934za,Tsai:1973py,ParticleDataGroup:2024cfk}. Ionized electrons contribute to the pair-production cross-section comparably to hydrogen, with the only difference being the energy threshold ($E_{th}=4m_e$ for electron targets, compared to the $E_{th}=2m_e$ for hydrogen).  Since most of the physics in our case plays out near the edge of the star, we can safely assume $Y_{p}\sim Y_{H^+}\sim Y_{e^-}$ to simplify our analysis (for details, see~\cite{Lu:2007wp}). These assumptions should hold to a great degree for stellar envelopes where hydrogen is the dominant species.

\textit{Positron Thermalization.}$-$ After their production, a fraction of the positrons lose energy rapidly, and annihilate at rest to produce the 511 keV line. Following \cite{Lu:2007wp}, we find that for the given condition of the stellar envelope, the thermalization rate is dominated by the excitation of the plasma electrons, using which we can calculate the average thermalization distance $d_{e^+\text{,therm}}(E_{e^+})$~\cite{RJGould:1972}. 
We also calculate that, for positron energies in the interval $10\text{ MeV}\leq E_{e^+} \leq 400\text{ MeV}$, the probability of annihilation at rest is $P_\text{ann}^{e^+}= 1- P_\text{In-Flight}(E_{e^+})\simeq 0.66$ - $0.91$, where $P_\text{In-Flight}$ is the in-flight annihilation probability (see supplemental material for details). For comparison, in the standard SN neutrino burst echo case, one gets  $P_\text{ann}^{e^+}\simeq 0.88$ \cite{Lu:2007wp,Lunardini:2023ilg}.

\textit{511 keV $\gamma$-ray Production.}$-$ After thermalization, the positron subsequently quickly undergoes annihilation. The dominant contribution to the 511 keV line comes from the  decay of the singlet positronium ($^1{\text{Ps}}$), which is formed with either free or bound electrons or from the conversion of the triplet state $^3{\text{Ps}}$~\cite{Lu:2007wp,Lunardini:2023ilg}.

For the 511 keV $\gamma$-rays produced in the annihilation, the mean free path is dominated by the Compton scattering; for this, the cross section is $\sigma_{\text{Compton}}^{511}\sim 3 \times 10^{-25}\text{ cm}^2$, and the corresponding mean free path is $\lambda_C \sim 10^{9}$ cm~\cite{Lunardini:2023ilg}. Therefore, the contributions to the 511 keV signal arise only from thermalized positron annihilations occuring in a thin layer of depth within $R_{\text{Compton}}\sim \lambda_C \sim 10^{9}$ cm underneath the stellar surface.  Note that, on average, only half of the gamma rays produced in this layer is detectable, as the rest propagates inward into the star and is absorbed. 

Upon collecting all the required ingredients, the differential 511 keV photon emission rate is described by the following expression (see  Fig.~\ref{fig:ALPcartoon} for a schematic representation): 
\begin{widetext}
\begin{align}
    \frac{d^3N_\gamma}{dt\,dE_a\,dr}(r, E_a, t) = &  \, 
    2\,\lambda_{a}^{-1}(r,E_a)  \frac{d^2N_a(r,E_a,t)}{dt\,dE_a}\Theta\left( t\right)\,\text{exp}\left[-\tau_{\text{a}}(R_\text{core},r,E_a)\right]\,f_{511}(r,E_a,m_a)\\ 
    f_{511}(r,E_a,m_a) = & \int_{r}^{R_\text{env}} dx\,\int_{E_{\gamma}^\text{min}}^{E_{\gamma}^\text{max}} dE_\gamma\,\int_{m_e}^{E_\gamma-m_e}  dE_{e^+}\lambda^{-1}_{\text{pair}}(x,E_\gamma)  \,\text{exp}\left[-\tau_{\text{pair}}(r,x,E_\gamma)\right]\,g_{{e^+}}(E_{e^+},E_\gamma)\,g_{{\gamma}}(E_\gamma,E_{a})\, \nonumber \\& \times P_\text{ann}^{e^+}(E_{e^+})\,\text{Box}\left( R_\text{env}-R_\text{compton} \leq x+d_{e^+\text{,therm}}(E_{e^+}) \leq R_\text{env}\right) 
    \label{eq:diffNgamma}
\end{align}
\end{widetext}
where $t$ is the elapsed time measured post-bounce, $r$ is the position at which an ALP with m.f.p. length $\lambda_{a}$ decays into two $\gamma$, and  
the quantities $\tau_a$ and $\tau_\text{pair}$ indicate optical depths ($\tau_i(r,R,E)\equiv\int_r^{R} \lambda^{-1}_i(r',E)\, \text{d}r'$). The function $f_{511}(r,E_a,m_a)$ provides the fraction of ALP decays leading to a 511 keV photon; it requires computing an integral over the pair-production distance $x$, where the m.f.p. $\lambda_{\text{pair}}$, positron $g_{{e^+}}(E_{e^+},E_\gamma)$ and photon energy distribution function $g_{{\gamma}}(E_\gamma,E_{a}$) should be included. Here $\text{Box}(a<z<b)$ is defined as $\text{Box}(z)=1$ for $a<z<b$ and $\text{Box}(z)=0$ elsewhere. It accounts for the limits on allowed region for positron thermalization. For details on the function $f_{511}(r,E_a,m_a)$, see supplemental material.

The expected gamma ray light curve at Earth can be calculated as follows
\begin{widetext}
\begin{equation}
     {\Phi_\gamma}(t_E)=  \frac{1}{4\pi D_{\text{SN}}^2} \int_{m_a}^{\infty} dE_a \int_0^1 d(\cos\theta)\int_{R_\text{core}}^{R_\text{env}} dr\,\frac{d^3N_\gamma}{dt\,dE_a\,dr}(t)\,; \quad \text{where }
    t=t_E-\bigg(\frac{E_a}{p_a}-1\bigg)r-R_{\text{env}}(1-\cos\theta)
    \label{eq:phigamma}
\end{equation}
\end{widetext}
where $D_{\text{SN}}$ is the distance to the SN. Note that $t_{E}$ refers to the time in a terrestrial detector's frame. It accounts for the lower velocities of ALPs with non-negligible masses (with respect to massless particles), and for the longer optical path traveled by photons away from the line of sight. Due to these effects, although the ALP flux produced in the supernova core is typically of duration $\mathcal{O}(10\ \text{s})$, the 511 keV photon flux at Earth exhibits an extended temporal tail. Note that $t=t_E$ for massless particles along the line of sight, and that $t=0$ is set as the time when the echo reaches Earth (i.e., the time of first detection of the accompanying neutrino burst).  In Table \ref{tab:intfluxdetails}, we give the time-integrated photon flux, $\phi_{511}=\int \Phi_\gamma(t_E)\,dt_E$, for three  representative points in the parameter space. We also show the time intervals which contain 90\% and 95\% of the total echo signal ($t_{90\%}$ and $t_{95\%}$ respectively). For fixed $m_a$ but smaller $G_{a\gamma}$ (compare case B and C), the decay length increases, implying that to generate the observed signal near the stellar surface (on scales of 
$R_{\text{env}}$ but within $R_{\text{compton}}$), lower energy decays dominate. However, because these ALPs are slower, the resulting signal is distributed over a longer timescale. For fixed $G_{a\gamma}$ but varying $m_a$ (compare case A and B), heavier ALPs exhibit shorter decay lifetimes, thereby shifting the dominant contribution to the signal toward faster ALPs corresponding to the signal arriving earlier than for the lighter $m_a$. We find that, for our reference value  $R_{env}=6\times10^{12}$ cm and $G_{a\gamma}$ sensitivity, the majority of the signal arrives within 1000 seconds. For SN1987A with a smaller stellar envelope ($R_\text{env}=2\times10^{12}$ cm), the corresponding arrival times (for $G_{a\gamma}$ it can probe) is found to be $t^{1987\text{A}}_{90\%} \sim 120$ s.  From the Table, it is also evident that the BSM echo signal will be significantly stronger than the one generated by the SN neutrino burst, which has an intensity of about $10^{-5}\text{ cm}^{-2}$ at 10 kpc~\cite{Lunardini:2023ilg}.  
\begin{table}[t]
    \centering
    \begin{tabular}{|c|c|c|c|}
    \hline
         ($G_{a\gamma}\,[\text{GeV}^{-1}],m_a$ [MeV])& $\phi_{511}$ (cm$^{-2}$)  & $t_{90\%}$ (s)& $t_{95\%}$ (s) \\ \hline 
         A : ($3\times10^{-12},\,50$)& $6.0\times10^{-2}$  & $\sim 1020$ & $\sim 1430$\\
        B : ($3\times10^{-12},300$)& $4.6\times10^{-2}$  &$\sim 690$ & $\sim 740$ \\
        C : ($1\times10^{-11},300$) & $1.5\times10^{-2}$ & $\sim 380$ & $\sim 410$ \\
      \hline   
    \end{tabular}
    \caption{The time-integrated 511 KeV gamma ray flux at Earth, assuming a SN  at $D_{\text{SN}}=10$ kpc with $R_{env}=6\times10^{12}$ cm, for representative values of $(G_{a\gamma},m_a)$. We also provide time durations corresponding to the arrival of $90\%$ and $95\%$ of the signal.}
    \label{tab:intfluxdetails}
\end{table}

\textit{Observations.}$-$ 
Although the progenitor for SN1987A was a blue supergiant, a compact H-rich stellar envelope of $5$-$15\text{ M}_\odot$ did exist~\cite{Woosley:1988,Utrobin:2005,Orlando:2025jzw}. During the SN1987A ($D_\text{SN}\simeq 51$ kpc), our predicted gamma-ray echo could have been detected by the Gamma Ray Burst detector aboard the Pioneer Venus Orbiter (PVO)~\cite{PVO:1979}, which was by coincidence pointing 
in the direction of the star~\cite{Jaffe:1995sw}. No excess over the background events was seen in any of the 4 channels, especially the $500-1000$ keV band, for $1500$ s prior and $8000$ s after SN1987A~\cite{Jaffe:1995sw}. Since the detector response function are not available, we try to estimate the background fluence measured by the detector. In Refs.~\cite{Jaffe:1995sw,Fenimore:2023blu}, the average background count rate in PVO $500-1000$ keV band is given as $\dot N_\text{bckg}\simeq 10~{\rm s^{-1}}$. Requiring a $3\sigma$ fluctuation of the background over $t^{1987\text{A}}_{90\%}$, we can set upper bound on the number of events from the gamma-ray echo  at $N_\text{max}=3\sqrt{\dot N_\text{bckg} t^{1987\text{A}}_{90\%}}= 104$ events. The effective area of PVO (over all 4-bands) is $\sim 22$ cm$^2$ \cite{PVO:Crider}, which is close to the detector's geometric area itself. Folding in with the  photon interaction probability in CsI~\cite{NIST:photon}, we obtain the effective area at 511 keV for overhead observation, $A_\text{eff}\simeq 17\text{ cm}^2$. We then get an upper bound on the time-integrated photon flux: $\phi_{511}\lesssim N_\text{max}/A_\text{eff}\simeq 6.18 ~\mathrm{ cm^{-2}}$.

Looking ahead at future galactic SN observations (representative distance $D_{\text{SN}}=10$ kpc), we consider the potential of the upcoming COSI \cite{Tomsick:2021wed,Lunardini:2023ilg}, which has $A_{\text{eff}}\simeq 20 {\text{\, cm}}^{2} $ at 511 keV, a wide field of view ($\sim 25\%$ of the sky) and an angular resolution better than $4.1^\circ$. As a representative of the next generation of detectors, we also consider 
AMEGO \cite{Kierans:2020otl,Lunardini:2023ilg}, which will have $A_\text{eff}\simeq 3 \times 10^{3} {\text{\, cm}}^{2} $ at 511 keV with an angular resolution of $\simeq 3^\circ$. Therefore, the magnitude of time-integrated echo flux 
required to observe $N=1$ event at COSI and AMEGO corresponds to  $1/A_\text{eff}\sim 5\times10^{-2}\text{ cm}^{-2}$ and $3.33\times10^{-4}\text{ cm}^{-2}$ respectively.  

The most relevant background in our case will arise from the diffuse galactic background. Assuming the galactic SN to be away from the galactic center, we take $d\phi_{\text{gal}}/d\Omega\simeq 10^{-4} \text{ cm}^{-2}\text{ s}^{-1}\text{ sr}^{-1}$ (for galactic latitude $b=0^\circ$ and longitude $l=60^\circ$)~\cite{Skinner2015}.
The background flux rate $\phi_{\text{gal}}$ for COSI and AMEGO will be $1.6 \times 10^{-6} \text{ cm}^{-2}\text{ s}^{-1}$ and $8.6 \times 10^{-7} \text{ cm}^{-2}\text{ s}^{-1}$,  with background number of events 
$N_\text{bckg}\simeq \phi_{\text{gal}} t_{\text{obs}} A_\text{eff} \simeq 0.03$ and $2.6$ respectively, for a signal integration time  $t_{\text{obs}}=1000$ s. These numbers imply that a measurement for COSI is essentially background-less, therefore we can set constraints by requiring a single event in the detector, and the flux sensitivity estimate given above holds. The future sensitivity for AMEGO to the $G_{a\gamma}$-$m_a$ parameter space at the $3\sigma$-level can be set by requiring a statistically significant excess over the background (using the log-likelihood method), which is at around $N=8$ events, corresponding to a signal flux of $2.72\times10^{-3}\text{ cm}^{-2}$. 

\textit{Results \& Discussion.}$-$ We finally show our results in Fig.~\ref{fig:MainConstraint}. The red region denotes the estimated 511 keV bound from PVO observations. The dashed blue and green curves denote the future exclusion sensitivity for the COSI and AMEGO detectors, respectively. We also include the relevant existing constraints in the $G_{a\gamma}$-$m_a$ parameter space from non-observation of high-energy $\gamma$-rays from SN1987A~\cite{Jaeckel:2017tud,Hoof:2022xbe,Muller:2023vjm} and binary neutron star merger event (GW170817)~\cite{Dev:2023hax,Diamond:2023cto}, shown as shaded gray regions. The recently derived constraint from observations of radiative decays in Type Ic SN has not been shown (with a significance of less than $2\sigma$)~\cite{Candon:2025ypl}. The new PVO bound derived in this work, reaching $G_{a\gamma}$ as low as $4\times10^{-11}\text{ GeV}^{-1}$, establishes the potential of the 511 keV signal as a unique probe. Although not as constraining as the SN1987A high-energy gamma-ray signal from decays outside the envelope~\cite{Jaeckel:2017tud,Hoof:2022xbe,Muller:2023vjm}, it outlines the complementarity of both signals for establishing ALP discovery in future. 

We observe that for the case of future galactic SN at $10$ kpc, the sensitivity with even a modest effective area $\sim 20\text{ cm}^{2}$ can probe ALP masses upto $~315$ MeV and coupling strengths as low as $1.7 \times 10^{-12}\text{ GeV}^{-1}$. AMEGO's larger $A_{\text{eff}}$ can extend the reach upto $m_a\sim 410$ MeV and $G_{a\gamma}\sim 6.4 \times 10^{-13}\text{ GeV}^{-1}$.

As stressed earlier, such a 511 keV echo signal should be present for all particles decaying radiatively, or to positrons directly, or to $\bar{\nu}_e$ initiating IBD. Our study and results can be extended to these other BSM species. We also note that to estimate the 511-keV bound for other cases beyond ALPs, instead of full electromagnetic cascade analysis with energy distribution taken into account, assuming energy equipartition in decays and for pair-production yields a good first-order estimate. 
Our work can be generalized to the case of carbon envelopes (hydrogen-stripped stars); this case will require using  the screened-regime cross-section for pair production.

We also point out that accompanying diffuse emission from in-flight annihilation as well as electromagnetic cascades from electron and photon energy losses near the stellar surface, will contribute to the high-energy gamma spectrum, thereby complicating the current analysis of the high-energy gamma-ray bounds obtained from SMM during SN1987A in some part of the parameter space. 
We expect that including these channels will lead to higher sensitivity to the echo, and therefore stronger bounds on the parameters. Therefore the exclusion/sensitivity regions in Fig. \ref{fig:MainConstraint} are robust. 

\textit{Acknowledgements.}$-$
We thank Giuseppe Lucente, Alexander Friedland, Flip Tanedo and Gustavo Marques- Tavares for insightful discussions. The work of GC and CL is supported by the NSF Award Number 2309973. GC also acknowledges the Fermilab Theory Group, the Center for Theoretical Underground Physics and Related Areas
(CETUP* 2025) and the Institute for Underground Science at SURF for hospitality and for providing a stimulating environment, where a part of this work was done.

\bibliography{ref}

@article{Bollig:2020xdr,
    author = "Bollig, Robert and DeRocco, William and Graham, Peter W. and Janka, Hans-Thomas",
    title = "{Muons in Supernovae: Implications for the Axion-Muon Coupling}",
    eprint = "2005.07141",
    archivePrefix = "arXiv",
    primaryClass = "hep-ph",
    doi = "10.1103/PhysRevLett.125.051104",
    journal = "Phys. Rev. Lett.",
    volume = "125",
    number = "5",
    pages = "051104",
    year = "2020",
    note = "[Erratum: Phys.Rev.Lett. 126, 189901 (2021)]"
}

@misc{SNprofile,
  title = "{Garching Core-collapse Supernova Research Archive}",
  howpublished = {\url{https://wwwmpa.mpa-garching.mpg.de/ccsnarchive/}},
  note = {}
}

@article{Mirizzi:2015eza,
    author = "Mirizzi, Alessandro and Tamborra, Irene and Janka, Hans-Thomas and Saviano, Ninetta and Scholberg, Kate and Bollig, Robert and Hudepohl, Lorenz and Chakraborty, Sovan",
    title = "{Supernova Neutrinos: Production, Oscillations and Detection}",
    eprint = "1508.00785",
    archivePrefix = "arXiv",
    primaryClass = "astro-ph.HE",
    doi = "10.1393/ncr/i2016-10120-8",
    journal = "Riv. Nuovo Cim.",
    volume = "39",
    number = "1-2",
    pages = "1--112",
    year = "2016"
}

@ARTICLE{BisnovatyiKogan:1975,
       author = {{Bisnovatyi-Kogan}, G.~S. and {Imshennik}, V.~S. and {Nadyozhin}, D.~K. and {Chechetkin}, V.~M.},
        title = "{Pulsed Gamma-Ray Emission from Neutron and Collapsing Stars and Supernovae}",
      journal = {Astrophys. Space Sci.},
         year = 1975,
        month = jun,
       volume = {35},
       number = {1},
        pages = {23-41},
          doi = {10.1007/BF00644821},
       adsurl = {https://ui.adsabs.harvard.edu/abs/1975Ap&SS..35...23B}
}

@ARTICLE{Ryazhskaya:1999,
       author = {{Ryazhskaya}, O.~G.},
        title = "{Presupernova {\ensuremath{\gamma}}-burst and some consequences of {\ensuremath{\nu}}{\ensuremath{\sim}}$_{e}$p-reaction.}",
      journal = {Nuovo Cimento C Geophysics Space Physics C},
         year = 1999,
        month = feb,
       volume = {22C},
        pages = {115-120},
       adsurl = {https://ui.adsabs.harvard.edu/abs/1999NCimC..22..115R}
}

@article{Lunardini:2023ilg,
    author = "Lunardini, Cecilia and Loeffler, Joshua and Mukhopadhyay, Mainak and Hurley, Matthew J. and Farag, Ebraheem and Timmes, F. X.",
    title = "{Photons from Neutrinos: The Gamma-Ray Echo of a Supernova Neutrino Burst}",
    eprint = "2312.13197",
    archivePrefix = "arXiv",
    primaryClass = "astro-ph.HE",
    doi = "10.3847/1538-4357/ad4546",
    journal = "Astrophys. J.",
    volume = "969",
    number = "2",
    pages = "149",
    year = "2024"
}

@article{Lu:2007wp,
    author = "Lu, Yu and Qian, Yong-Zhong",
    title = "{Neutrino-Induced Gamma-Ray Emission from Supernovae}",
    eprint = "0709.0501",
    archivePrefix = "arXiv",
    primaryClass = "astro-ph",
    doi = "10.1103/PhysRevD.76.103002",
    journal = "Phys. Rev. D",
    volume = "76",
    pages = "103002",
    year = "2007"
}

@book{Raffelt:1996wa,
    author = "Raffelt, G. G.",
    title = "{Stars as laboratories for fundamental physics}: {The astrophysics of neutrinos, axions, and other weakly interacting particles}",
    isbn = "978-0-226-70272-8",
    month = "5",
    year = "1996"
}

@article{Caputo:2022mah,
    author = "Caputo, Andrea and Janka, Hans-Thomas and Raffelt, Georg and Vitagliano, Edoardo",
    title = "{Low-Energy Supernovae Severely Constrain Radiative Particle Decays}",
    eprint = "2201.09890",
    archivePrefix = "arXiv",
    primaryClass = "astro-ph.HE",
    doi = "10.1103/PhysRevLett.128.221103",
    journal = "Phys. Rev. Lett.",
    volume = "128",
    number = "22",
    pages = "221103",
    year = "2022"
}

@article{Oberauer:1993yr,
    author = "Oberauer, L. and Hagner, C. and Raffelt, G. and Rieger, E.",
    title = "{Supernova bounds on neutrino radiative decays}",
    doi = "10.1016/0927-6505(93)90004-W",
    journal = "Astropart. Phys.",
    volume = "1",
    pages = "377--386",
    year = "1993"
}

@article{ParticleDataGroup:2024cfk,
    author = "Navas, S. and others",
    collaboration = "Particle Data Group",
    title = "{Review of particle physics}",
    doi = "10.1103/PhysRevD.110.030001",
    journal = "Phys. Rev. D",
    volume = "110",
    number = "3",
    pages = "030001",
    year = "2024"
}

@article{Tsai:1973py,
    author = "Tsai, Yung-Su",
    title = "{Pair Production and Bremsstrahlung of Charged Leptons}",
    reportNumber = "SLAC-PUB-1365",
    doi = "10.1103/RevModPhys.46.815",
    journal = "Rev. Mod. Phys.",
    volume = "46",
    pages = "815",
    year = "1974",
    note = "[Erratum: Rev.Mod.Phys. 49, 421--423 (1977)]"
}

@article{Bethe:1934za,
    author = "Bethe, H. and Heitler, W.",
    title = "{On the Stopping of fast particles and on the creation of positive electrons}",
    doi = "10.1098/rspa.1934.0140",
    journal = "Proc. Roy. Soc. Lond. A",
    volume = "146",
    pages = "83--112",
    year = "1934"
}

@ARTICLE{PVO:1979,
       author = {{Evans}, W.~D. and {Glore}, J.~P. and {Klebesadel}, R.~W. and {Laros}, J.~G. and {Tech}, E.~R. and {Spalding}, R.~E.},
        title = "{Gamma-Ray Burst Observations by Pioneer Venus Orbiter}",
      journal = {Science},
         year = 1979,
        month = jul,
       volume = {205},
       number = {4401},
        pages = {119-121},
          doi = {10.1126/science.205.4401.119},
       adsurl = {https://ui.adsabs.harvard.edu/abs/1979Sci...205..119E}
}

@article{Jaffe:1995sw,
    author = "Jaffe, Andrew H. and Turner, Michael S.",
    title = "{Gamma-rays and the decay of neutrinos from SN1987A}",
    eprint = "astro-ph/9601104",
    archivePrefix = "arXiv",
    reportNumber = "FERMILAB-PUB-95-397-A, CITA-95-26",
    doi = "10.1103/PhysRevD.55.7951",
    journal = "Phys. Rev. D",
    volume = "55",
    pages = "7951--7959",
    year = "1997"
}

@article{PVO:Crider,
    author = {Crider, A. and Fenimore, E. E.},
    title = {A search for March 5th-like bursts in the PVO database },
    journal = {AIP Conference Proceedings},
    volume = {384},
    number = {1},
    pages = {926-930},
    year = {1996},
    month = {08},
    issn = {0094-243X},
    doi = {10.1063/1.51619},
    url = {https://doi.org/10.1063/1.51619},
}

@article{Fenimore:2023blu,
    author = "Fenimore, E. E. and Crider, A. and Zand, J. J. M. int and Klebesadel, R. W. and Laros, J. G. and Meier, M.",
    title = "{The Pioneer Venus Orbiter Catalog of Gamma-Ray Bursts}",
    eprint = "2302.12859",
    archivePrefix = "arXiv",
    primaryClass = "astro-ph.HE",
    reportNumber = "LA-UR 23-21872",
    month = "2",
    year = "2023"
}

@article{Skinner2015,
  author  = {Gerald K. Skinner and Roland Diehl and Xiaoling Zhang and Laurent Bouchet and Pierre Jean},
  title   = {The Galactic distribution of the 511 keV e$^{+}$/e$^{-}$ annihilation radiation},
  journal = {Proceedings of Science (PoS)},
  volume  = {228},
  pages   = {054},
  year    = {2015},
  month   = mar,
  doi     = {10.22323/1.228.0054},
  url     = {https://pos.sissa.it/228/054},
  note    = {Proceedings of the 10th INTEGRAL Workshop (Integral2014), Annapolis, MD, USA}
}

@article{Kamiokande-II:1987idp,
    author = "Hirata, K. and others",
    editor = "Wali, K. C.",
    collaboration = "Kamiokande-II",
    title = "{Observation of a Neutrino Burst from the Supernova SN 1987a}",
    reportNumber = "UT-ICEPP-87-01, UPR-142E",
    doi = "10.1103/PhysRevLett.58.1490",
    journal = "Phys. Rev. Lett.",
    volume = "58",
    pages = "1490--1493",
    year = "1987"
}

@article{Bionta:1987qt,
    author = "Bionta, R. M. and others",
    title = "{Observation of a Neutrino Burst in Coincidence with Supernova SN 1987a in the Large Magellanic Cloud}",
    reportNumber = "UCI-NEUTRINO-87-10",
    doi = "10.1103/PhysRevLett.58.1494",
    journal = "Phys. Rev. Lett.",
    volume = "58",
    pages = "1494",
    year = "1987"
}

@article{Raffelt:1990yz,
    author = "Raffelt, Georg G.",
    title = "{Astrophysical methods to constrain axions and other novel particle phenomena}",
    reportNumber = "MPI-PAE-PTH-29-90",
    doi = "10.1016/0370-1573(90)90054-6",
    journal = "Phys. Rept.",
    volume = "198",
    pages = "1--113",
    year = "1990"
}

@article{Choi:1987sd,
    author = "Choi, Kiwoon and Kim, C. W. and Kim, Jewan and Lam, W. P.",
    title = "{Constraints on the Majoron Interactions From the Supernova {SN1987A}}",
    reportNumber = "JHU-TIPAC-8722",
    doi = "10.1103/PhysRevD.37.3225",
    journal = "Phys. Rev. D",
    volume = "37",
    pages = "3225",
    year = "1988"
}

@article{Kolb:1988pe,
    author = "Kolb, Edward W. and Turner, Michael S.",
    title = "{Limits to the Radiative Decays of Neutrinos and Axions from Gamma-Ray Observations of SN 1987a}",
    reportNumber = "FERMILAB-PUB-87-223-A, FERMILAB-PUB-87-223-A-REV",
    doi = "10.1103/PhysRevLett.62.509",
    journal = "Phys. Rev. Lett.",
    volume = "62",
    pages = "509",
    year = "1989"
}

@article{DeRocco:2019njg,
    author = "DeRocco, William and Graham, Peter W. and Kasen, Daniel and Marques-Tavares, Gustavo and Rajendran, Surjeet",
    title = "{Observable signatures of dark photons from supernovae}",
    eprint = "1901.08596",
    archivePrefix = "arXiv",
    primaryClass = "hep-ph",
    doi = "10.1007/JHEP02(2019)171",
    journal = "JHEP",
    volume = "02",
    pages = "171",
    year = "2019"
}

@article{Calore:2021klc,
    author = "Calore, Francesca and Carenza, Pierluca and Giannotti, Maurizio and Jaeckel, Joerg and Lucente, Giuseppe and Mirizzi, Alessandro",
    title = "{Supernova bounds on axionlike particles coupled with nucleons and electrons}",
    eprint = "2107.02186",
    archivePrefix = "arXiv",
    primaryClass = "hep-ph",
    doi = "10.1103/PhysRevD.104.043016",
    journal = "Phys. Rev. D",
    volume = "104",
    number = "4",
    pages = "043016",
    year = "2021"
}

@article{Calore:2021lih,
    author = "Calore, Francesca and Carenza, Pierluca and Giannotti, Maurizio and Jaeckel, Joerg and Lucente, Giuseppe and Mastrototaro, Leonardo and Mirizzi, Alessandro",
    title = "{511~keV line constraints on feebly interacting particles from supernovae}",
    eprint = "2112.08382",
    archivePrefix = "arXiv",
    primaryClass = "hep-ph",
    doi = "10.1103/PhysRevD.105.063026",
    journal = "Phys. Rev. D",
    volume = "105",
    number = "6",
    pages = "063026",
    year = "2022"
}

@article{Dev:2023hax,
    author = "Dev, P. S. Bhupal and Fortin, Jean-Fran{\c{c}}ois and Harris, Steven P. and Sinha, Kuver and Zhang, Yongchao",
    title = "{First Constraints on the Photon Coupling of Axionlike Particles from Multimessenger Studies of the Neutron Star Merger GW170817}",
    eprint = "2305.01002",
    archivePrefix = "arXiv",
    primaryClass = "hep-ph",
    reportNumber = "INT-PUB-23-014",
    doi = "10.1103/PhysRevLett.132.101003",
    journal = "Phys. Rev. Lett.",
    volume = "132",
    number = "10",
    pages = "101003",
    year = "2024"
}

@article{Diamond:2023cto,
    author = "Diamond, Melissa and Fiorillo, Damiano F. G. and Marques-Tavares, Gustavo and Tamborra, Irene and Vitagliano, Edoardo",
    title = "{Multimessenger Constraints on Radiatively Decaying Axions from GW170817}",
    eprint = "2305.10327",
    archivePrefix = "arXiv",
    primaryClass = "hep-ph",
    doi = "10.1103/PhysRevLett.132.101004",
    journal = "Phys. Rev. Lett.",
    volume = "132",
    number = "10",
    pages = "101004",
    year = "2024"
}

@article{Muller:2023vjm,
    author = {M{\"u}ller, Eike and Calore, Francesca and Carenza, Pierluca and Eckner, Christopher and Marsh, M. C. David},
    title = "{Investigating the gamma-ray burst from decaying MeV-scale axion-like particles produced in supernova explosions}",
    eprint = "2304.01060",
    archivePrefix = "arXiv",
    primaryClass = "astro-ph.HE",
    doi = "10.1088/1475-7516/2023/07/056",
    journal = "JCAP",
    volume = "07",
    pages = "056",
    year = "2023"
}

@ARTICLE{Woosley:1988,
       author = {{Woosley}, S.~E. and {Pinto}, Philip A. and {Ensman}, L.},
        title = "{Supernova 1987A: Six Weeks Later}",
      journal = {\apj},
         year = 1988,
        month = jan,
       volume = {324},
        pages = {466},
          doi = {10.1086/165908},
       adsurl = {https://ui.adsabs.harvard.edu/abs/1988ApJ...324..466W}
}

@ARTICLE{Utrobin:2005,
       author = {{Utrobin}, V.~P.},
        title = "{Supernova 1987A: The Envelope Mass and the Explosion Energy}",
      journal = {Astron. Lett.},
         year = 2005,
        month = dec,
       volume = {31},
       number = {12},
        pages = {806-815},
          doi = {10.1134/1.2138767},
       adsurl = {https://ui.adsabs.harvard.edu/abs/2005AstL...31..806U}
}

@article{Orlando:2025jzw,
    author = "Orlando, S. and others",
    title = "{Tracing the ejecta structure of supernova 1987A: Insights and diagnostics from 3D magnetohydrodynamic simulations}",
    eprint = "2504.19896",
    archivePrefix = "arXiv",
    primaryClass = "astro-ph.HE",
    doi = "10.1051/0004-6361/202554862",
    journal = "Astron. Astrophys.",
    volume = "699",
    pages = "A305",
    year = "2025"
}

@article{Tomsick:2021wed,
    author = "Tomsick, John A.",
    collaboration = "COSI",
    title = "{The Compton Spectrometer and Imager Project for MeV Astronomy}",
    eprint = "2109.10403",
    archivePrefix = "arXiv",
    primaryClass = "astro-ph.IM",
    doi = "10.22323/1.395.0652",
    journal = "PoS",
    volume = "ICRC2021",
    pages = "652",
    year = "2021"
}

@article{e-ASTROGAM:2017pxr,
    author = "Tavani, M. and others",
    editor = "De Angelis, A. and Tatischeff, V. and Grenier, I. A. and McEnery, J. and Mallamaci, M.",
    collaboration = "e-ASTROGAM",
    title = "{Science with e-ASTROGAM: A space mission for MeV{\textendash}GeV gamma-ray astrophysics}",
    eprint = "1711.01265",
    archivePrefix = "arXiv",
    primaryClass = "astro-ph.HE",
    doi = "10.1016/j.jheap.2018.07.001",
    journal = "JHEAp",
    volume = "19",
    pages = "1--106",
    year = "2018"
}

@article{Kierans:2020otl,
    author = "Kierans, Carolyn A.",
    collaboration = "AMEGO Team",
    title = "{AMEGO: Exploring the Extreme Multimessenger Universe}",
    eprint = "2101.03105",
    archivePrefix = "arXiv",
    primaryClass = "astro-ph.IM",
    doi = "10.1117/12.2562352",
    journal = "Proc. SPIE Int. Soc. Opt. Eng.",
    volume = "11444",
    pages = "1144431",
    year = "2020"
}

@article{Raffelt:1985nk,
    author = "Raffelt, Georg G.",
    title = "{Astrophysical Axion Bounds Diminished by Screening Effects}",
    reportNumber = "MPI-PAE-PTH-51-85",
    doi = "10.1103/PhysRevD.33.897",
    journal = "Phys. Rev. D",
    volume = "33",
    pages = "897",
    year = "1986"
}

@article{DiLella:2000dn,
    author = "Di Lella, L. and Pilaftsis, A. and Raffelt, G. and Zioutas, K.",
    title = "{Search for solar Kaluza-Klein axions in theories of low scale quantum gravity}",
    eprint = "hep-ph/0006327",
    archivePrefix = "arXiv",
    reportNumber = "WUE-ITP-2000-015",
    doi = "10.1103/PhysRevD.62.125011",
    journal = "Phys. Rev. D",
    volume = "62",
    pages = "125011",
    year = "2000"
}

@article{Ferreira:2022xlw,
    author = {Ferreira, Ricardo Z. and Marsh, M. C. David and M{\"u}ller, Eike},
    title = "{Strong supernovae bounds on ALPs from quantum loops}",
    eprint = "2205.07896",
    archivePrefix = "arXiv",
    primaryClass = "hep-ph",
    doi = "10.1088/1475-7516/2022/11/057",
    journal = "JCAP",
    volume = "11",
    pages = "057",
    year = "2022"
}

@misc{NIST:photon,
 author={},
  title = "{NIST XCOM : Photon Cross Sections Database}",
  howpublished = {\url{https://physics.nist.gov/PhysRefData/Xcom/html/xcom1.html}},
  note = {}
}

@article{Falk:1978kf,
    author = "Falk, Sydney W. and Schramm, David N.",
    title = "{Limits From Supernovae on Neutrino Radiative Lifetimes}",
    reportNumber = "EFI-78-35-CHICAGO",
    doi = "10.1016/0370-2693(78)90417-3",
    journal = "Phys. Lett. B",
    volume = "79",
    pages = "511",
    year = "1978"
}

@article{Chupp:1989kx,
    author = "Chupp, E. L. and Vestrand, W. T. and Reppin, C.",
    title = "{Experimental Limits on the Radiative Decay of {SN1987A} Neutrinos}",
    doi = "10.1103/PhysRevLett.62.505",
    journal = "Phys. Rev. Lett.",
    volume = "62",
    pages = "505--508",
    year = "1989"
}

@article{Jaeckel:2017tud,
    author = "Jaeckel, J. and Malta, P. C. and Redondo, J.",
    title = "{Decay photons from the axionlike particles burst of type II supernovae}",
    eprint = "1702.02964",
    archivePrefix = "arXiv",
    primaryClass = "hep-ph",
    doi = "10.1103/PhysRevD.98.055032",
    journal = "Phys. Rev. D",
    volume = "98",
    number = "5",
    pages = "055032",
    year = "2018"
}

@article{Lucente:2020whw,
    author = "Lucente, Giuseppe and Carenza, Pierluca and Fischer, Tobias and Giannotti, Maurizio and Mirizzi, Alessandro",
    title = "{Heavy axion-like particles and core-collapse supernovae: constraints and impact on the explosion mechanism}",
    eprint = "2008.04918",
    archivePrefix = "arXiv",
    primaryClass = "hep-ph",
    doi = "10.1088/1475-7516/2020/12/008",
    journal = "JCAP",
    volume = "12",
    pages = "008",
    year = "2020"
}

@article{Dicus:1979ch,
    author = "Dicus, Duane A. and Kolb, Edward W. and Teplitz, Vigdor L. and Wagoner, Robert V.",
    title = "{Astrophysical Bounds on Very Low Mass Axions}",
    reportNumber = "Print-80-0024 (CAL TECH), OAP-582",
    doi = "10.1103/PhysRevD.22.839",
    journal = "Phys. Rev. D",
    volume = "22",
    pages = "839",
    year = "1980"
}

@article{Hoof:2022xbe,
    author = "Hoof, Sebastian and Schulz, Lena",
    title = "{Updated constraints on axion-like particles from temporal information in supernova SN1987A gamma-ray data}",
    eprint = "2212.09764",
    archivePrefix = "arXiv",
    primaryClass = "hep-ph",
    reportNumber = "TTP22-072",
    doi = "10.1088/1475-7516/2023/03/054",
    journal = "JCAP",
    volume = "03",
    pages = "054",
    year = "2023"
}

@article{Candon:2025ypl,
    author = "Cand{\'o}n, Francisco R. and Fiorillo, Damiano F. G. and Janka, Hans-Thomas and van Baal, Bart F. A. and Vitagliano, Edoardo",
    title = "{Small Progenitors, Large Couplings: Type Ic Supernova Constraints on Radiatively Decaying Particles}",
    eprint = "2509.18253",
    archivePrefix = "arXiv",
    primaryClass = "hep-ph",
    month = "9",
    year = "2025"
}

@article{Diamond:2023scc,
    author = "Diamond, Melissa and Fiorillo, Damiano F. G. and Marques-Tavares, Gustavo and Vitagliano, Edoardo",
    title = "{Axion-sourced fireballs from supernovae}",
    eprint = "2303.11395",
    archivePrefix = "arXiv",
    primaryClass = "hep-ph",
    doi = "10.1103/PhysRevD.107.103029",
    journal = "Phys. Rev. D",
    volume = "107",
    number = "10",
    pages = "103029",
    year = "2023",
    note = "[Erratum: Phys.Rev.D 108, 049902 (2023)]"
}

@ARTICLE{RJGould:1972,
       author = {{Gould}, R.~J.},
        title = "{Energy loss of fast electrons and positrons in a plasma}",
      journal = {Physica},
         year = 1972,
        month = jul,
       volume = {60},
       number = {1},
        pages = {145-154},
          doi = {10.1016/0031-8914(72)90227-3},
       adsurl = {https://ui.adsabs.harvard.edu/abs/1972Phy....60..145G}
}

@ARTICLE{RJGould:1989,
       author = {{Gould}, Robert J.},
        title = "{Direct Positron Annihilation and Positronium Formation in Thermal Plasmas}",
      journal = {\apj},
         year = 1989,
        month = sep,
       volume = {344},
        pages = {232},
          doi = {10.1086/167792},
       adsurl = {https://ui.adsabs.harvard.edu/abs/1989ApJ...344..232G}
}

\clearpage
\appendix

\onecolumngrid
\begin{center}
    \textbf{\Large Supplemental Material}
\end{center}
\begin{center}
    \textbf{\large 511 keV Gamma Ray Echo from Particle Decays in Supernovae}
\end{center}
In this supplemental material, we provide the relevant details for the ALP production rates, pair-production cross-section, positron thermalization and PVO effective area.
\newline

\twocolumngrid
\section{ALP Production Rates}
\label{app:ALPProd}
\begin{figure}[t!]
    \centering
    \includegraphics[width=\linewidth]{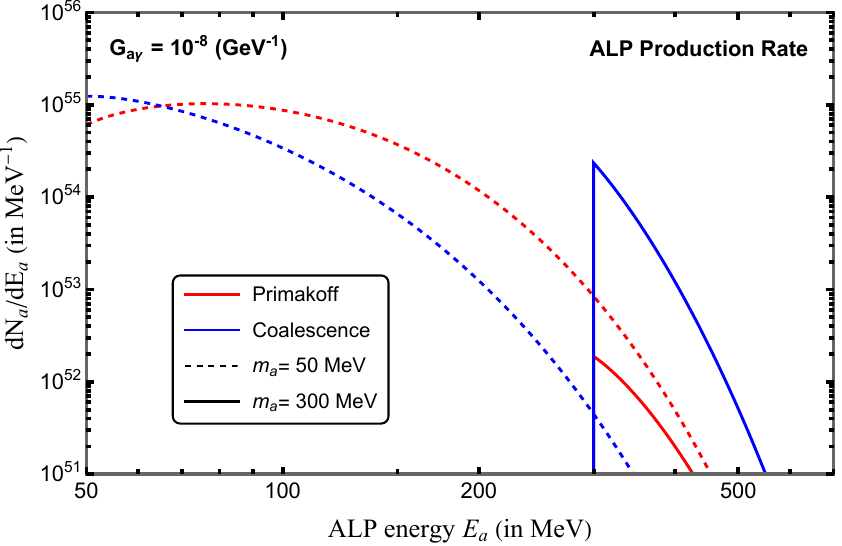}
    \caption{Time-integrated ALP production rate $dN_a/dE_a$ as function of ALP energy $E_a$ through Primakoff scattering and Coalescence process for ALP masses $m_a=(50,300)$ MeV, for a reference $G_{a\gamma}=10^{-8}\text{ GeV}^{-1}$.}
    \label{fig:ALPProd}
\end{figure}
The primary expressions for ALP production rate $\frac{d^2N_a(r,E_a,t)}{dt\,dE_a}$ from Primakoff and Coalescence process have been provided in Eqs.~\eqref{eq:sub1},\eqref{eq:sub2},\eqref{eq:main1}. The Primakoff rate depends on $f_\text{P}$ given by~\cite{Raffelt:1985nk,DiLella:2000dn}
\begin{align}
f_\text{P}= &
\frac{
\left[(k+p)^2 + k_s^2\right]\left[(k-p)^2 + k_s^2\right]
}{
16 k_s^2 k p
}
\log\!\left(
\frac{(k+p)^2 + k_s^2}{(k-p)^2 + k_s^2}
\right) \nonumber\\
&-
\frac{(k^2 - p^2)^2}{16 k_s^2 k p}
\log\!\left(
\frac{(k+p)^2}{(k-p)^2}
\right)
-\frac{1}{4}.
\end{align}
where $p=\sqrt{E_a^2-m_a^2}$, $k=\sqrt{E_a^2-\omega_P^2}$, relativistic electron plasma frequency $\omega_P=(4\alpha/3\pi)(\mu_e^2+\pi^2T^2/3)$ with $\mu_e$ being the electron chemical potential, and $k_s$ is the screening scale 
\begin{equation}
    k_s^2= \frac{4\pi\alpha}{T}\hat{n}
\end{equation}
where $\hat{n}\simeq (1-Y_n)n_B$ is the effective charged particle density, $Y_n$ is the neutron fraction and $n_B$ is the total baryon number density. 

The averaged photon stimulation factor $f_\text{B}$ required for Coalescence rate is~\cite{Caputo:2022mah,Ferreira:2022xlw}
\begin{equation}
\begin{aligned}
f_B &= \frac{1}{1 - e^{-\omega / T}} 
\frac{2T}{p} 
\log \left[
\frac{
e^\frac{\omega+p }{4T} - e^{-\frac{\omega+p }{4T}}
}{
e^\frac{\omega-p }{4T} - e^{-\frac{\omega-p }{4T}}
}
\right]
\end{aligned}
\end{equation}
where $\omega=E_a$ and $p=\sqrt{E_a^2-m_a^2}$.

For reference, we provide the individual time-integrated ALP production rate contributions from Primakoff and Coalescence process for ALP masses of 50 MeV and 300 MeV, shown in Fig.~\ref{fig:ALPProd}. It can be clearly seen that while for 50 MeV ALP, both Primakoff and Coalescence contribute roughly equally -- the production rate for heavier ALP mass of 300 MeV is dominated by the coalescence rate. We also note that while our characteristic spectrum shapes matches with results in Ref.~\cite{Muller:2023vjm}, our production rate is consistently lower by roughly factor of $5$, which we attribute to the difference in the SN simulation profiles used in both works. 

\section{Energy and Angular Distribution }
\label{app:EnergyAngDist}
The ALP decay to two photons in its rest-frame is isotropic, implying the energy distribution in the SN frame to be a box function 
\begin{align}
    g_{{\gamma}}(E_\gamma,E_{a})=&\frac{1}{\gamma_a\,\beta_a\,m_a}\Theta\left[E_\gamma-\frac{\gamma_a(1-\beta_a)m_a}{2}\right]
    \nonumber \\ & \times \Theta\left[\frac{\gamma_a(1+\beta_a)m_a}{2}-E_\gamma\right]
\end{align}
where $E_{\gamma}^\text{min}$ and $E_{\gamma}^\text{max}$ as referred to in the Eq.~\eqref{eq:diffNgamma}, represent the minimum and maximum energies allowed by the distribution function.
The differential angular distribution is given by
\begin{equation}
  \frac{1}{\Gamma_a} \frac{d\Gamma_a}{d\cos\theta}
= \frac{1 - \beta_a^2}{2\,\bigl(1 - \beta_a\,\cos\theta\bigr)^{2}}
\end{equation}
clearly shows a forward peaked distribution near $\beta_a \simeq 1$. 

\section{Pair Production}
The combined general Bethe-Heitler differential pair-production cross-section is~\cite{Bethe:1934za,Tsai:1973py,ParticleDataGroup:2024cfk} 
\begin{align}
\frac{d\sigma_\text{pair}}{dx}
&= \frac{4 Z^2 \alpha r_e^2}{3}
   \Bigl[\,(1 - 4\,x(1-x))\,\Phi_1(\delta)
        + x(1-x)\,\Phi_2(\delta)\Bigr],
\notag\\
&\quad
x = \frac{E_+}{E_\gamma}, 
\quad 0 < x < 1.
\end{align}
where the $\Phi_i$ are defined as 
\begin{align}
\Phi_1(\delta)
&= \ln\!\biggl(\frac{B_1 + \delta}{\delta}\biggr)
  - \frac{B_1}{B_1 + \delta},
  \\[6pt]
\Phi_2(\delta)
&= \ln\!\biggl(\frac{B_2 + \delta}{\delta}\biggr)
  - \frac{2 B_2}{B_2 + \delta}
  + \frac{B_2^2}{2\,(B_2 + \delta)^2}.
\end{align}
with $B_1 = 184\,Z^{-1/3},
B_2 = 1194\,Z^{-2/3}$ and the screening parameter given by 
\begin{equation}
    \delta(x)
= \frac{(m_e c^2)^2}{E_\gamma^2}
  \,\frac{\hbar^2}{a^2}\,\frac{1}{x(1-x)},\,a = 0.8853\,a_0\,Z^{-1/3}.
\end{equation}
where $a_0$ is the Bohr radius.

The pair-production probability crucially depends on the Bethe-Heitler cross-section ($\sigma_{\mathrm{pair}}$) and the hydrogen number density ($n_{\text H}$).
$$
\lambda^{-1}_{\text{pair}}(E_\gamma; x) = n_{\text H}\,\sigma_{\mathrm{pair}}(E_\gamma).
$$
where the cross-section for pair-production in the Coulomb field of the proton ($Z$) (no screening case) is
\begin{equation}
   \sigma_{\mathrm{pair}}(E_\gamma)= \alpha\,Z^2\,r_e^2\bigg( \frac{28}{9}\log\frac{2E_\gamma}{m_e}-\frac{218}{27}\bigg)
\end{equation}
and for complete screening case is 
\begin{equation}
   \sigma_{\mathrm{pair}}(E_\gamma)= \alpha\,Z^2\,r_e^2\bigg( \frac{28}{9}\log\frac{183}{Z^{1/3}}-\frac{2}{27}\bigg)
\end{equation}
where $r_e$ is the classical electron radius, the screening scale $E_{\text{scr}}\sim m_e^2\,a_0\,Z^{-1/3}$. In a hydrogen envelope, $E_{\text{scr}}=62$ MeV. Since the hydrogen in our region of interest is almost completely ionized, we use the no screening case cross-section. While the angular distribution is peaked in the forward direction ($\theta\sim m_e/E_\gamma$), the energy distribution  has a distinctive plateau shape for the no-screening case while peaked more towards the endpoints for the complete screened case given by 
\begin{equation}
    g_{{e^+}}(E_{e^+},E_\gamma)=\frac{1}{\sigma_\text{pair}}\frac{d\sigma_\text{pair}}{dx}(x, E_\gamma)
\end{equation}
where $x = E_+/E_\gamma$.

Since pair-production can occur far away from the production point, we need to also account for the photon intensity attenuation, which can arise due to various other processes such as Compton scattering off atomic electrons. However for high-energy photons, Compton scattering is sharply forward-peaked leading to minimal energy loss. Therefore, the only attenuation in our case arises dominantly from the pair-production process.
\begin{equation}
     I(r+x) = I(r)\,\text{exp}\left[-\tau_{\text{pair}}(r,x,E_\gamma)\right]
\end{equation}
where $I(r)$ is the photon intensity at $r$.

\section{Positron Thermalization}
The thermalization rate is dominated by the excitation of the plasma electrons with the energy loss rate given by~\cite{RJGould:1972,Lu:2007wp}
\begin{equation}
    - \left( \frac{dE_{e^+}}{dx} \right)_{\text{ex,pl}} 
=   \frac{\omega_p^2\,e^2}{v^2}\left(  \ln \left[ \frac{ \sqrt{2\epsilon (\gamma - 1)} m_e v }{ \omega_p } \right] + b(\gamma, \epsilon) \right)
\end{equation}
In the above equations, $v$ is the positron velocity,  $\gamma = 1/\sqrt{1 - (v/c)^2}$, $\epsilon$ is the maximum fraction of the positron energy lost in a single interaction and is taken to be $1/2$, $\omega_p = \sqrt{4 \pi \rho Y_e N_A e^2 / m_e}$ is the plasma frequency, and

\begin{align}
b(\gamma, \epsilon) &= - \left( \frac{\gamma^2 - 1}{2\gamma^2} \right) \epsilon 
+ \frac{1}{8} \left( \frac{\gamma - 1}{\gamma} \right)^2 \epsilon^2 \notag \\
&\quad - \frac{1}{2} \left( \frac{\gamma - 1}{\gamma + 1} \right)
\left[ \left( \frac{\gamma + 2}{\gamma} \right) \epsilon 
- \frac{\gamma^2 - 1}{\gamma^2} \epsilon^2 \right] \notag \\
&\quad + \left( \frac{\gamma - 1}{\gamma} \right) \epsilon^3 
+ \left( \frac{\gamma - 1}{\gamma} \right)^2 \epsilon^2 \notag \\
&\quad \times \left[ \left( \frac{1}{2} + \frac{1}{\gamma} + \frac{3}{2 \gamma^2} \right) 
\frac{\epsilon^2}{2} - \left( \frac{\gamma - 1}{\gamma} \right)^2 
\left( \frac{\epsilon^3}{3} - \frac{\epsilon^4}{4} \right) \right]. 
\end{align}
The thermalization distance can then be calculated as follows~\cite{RJGould:1972,Lu:2007wp}
\begin{equation}
   d_{e^+\text{,therm}} = \int_{E_{e^+}}^{m_e}\frac{dE_{e^+}}{ \left( \frac{dE_{e^+}}{dx} \right)_{\text{ex,pl}} }
\end{equation}
where the upper limit for the $E_{e^+}$ integral represents the thermalized positron with minuscule thermal kinetic energy $\sim 2$ eV, with rest mass contribution included.  

However we also need to account for the in-flight annihilation case, where a relativistic positron may annihilate before complete thermalization can occur. The probability for in-flight annihilation can be calculated as follows~\cite{RJGould:1989,Lu:2007wp}
\begin{equation}
    P_\text{In-Flight}(E_{e^+}) = \rho Y_e N_A \int_{E_{e^+}}^{m_e}\frac{\sigma_{e^+e^-}^{\text{rel}}\,dE_{e^+}}{ \left( \frac{dE_{e^+}}{dx} \right)_{\text{ex,pl}} }
\end{equation}
where $\sigma_{e^+e^-}^{\text{rel}}$ is the relativistic $e^+e^-$ annihilation cross-section given by 
\begin{align}
\sigma_{e^+e^-}^{\text{rel}}
&= \left( \frac{e^2}{m_e} \right)^2 \frac{\pi}{\gamma + 1}
\left[
\frac{\gamma^2 + 4\gamma + 1}{\gamma^2 - 1} 
\ln \left( \gamma + \sqrt{\gamma^2 - 1} \right) \right. \notag \\
&\quad \left. - \frac{\gamma + 3}{\sqrt{\gamma^2 - 1}}
\right]
\end{align}

\section{$f_{511}$ function}
\label{app:frac511}
\begin{figure}[h!]
    \centering
    \includegraphics[width=\linewidth]{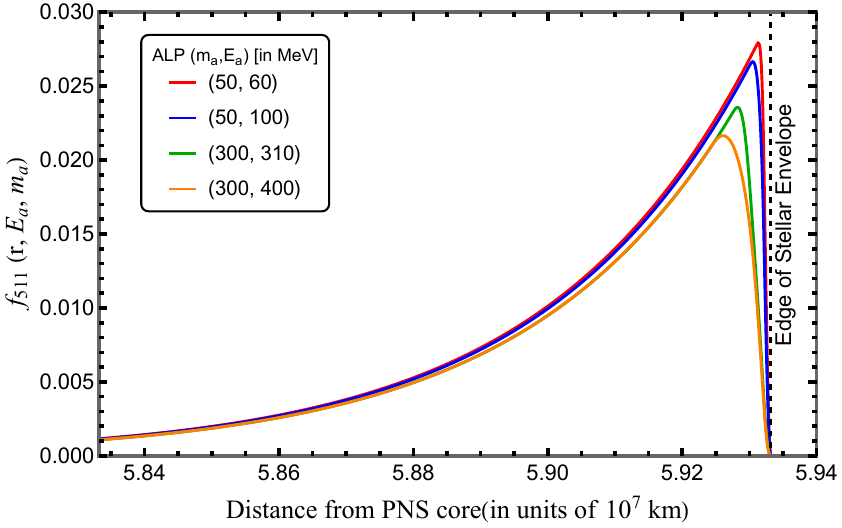}
    \caption{$f_{511}(r,E_a,m_a)$ as a function of distance from the PNS core ($r$) (defined in Eq.~\eqref{eq:diffNgamma}) for different values of ALP mass and energies.}
    \label{fig:frac511}
\end{figure}

We show the $f_{511}(r,E_a,m_a)$ function defined in Eq.~\eqref{eq:diffNgamma}, for different values of ALP mass and energies ($m_a=(50,300)$ MeV) as a function of distance from the PNS core ($r$) in Fig.~\ref{fig:frac511}. It can be seen clearly that the dominant contribution to the 511 keV signal from ALP decays arises at most $\sim 10^{11}$ cm from near the stellar surface. 

\section{PVO Effective Area}
\label{app:PVO}
In absence of the PVO's detector response function, we estimate the effective area at $E_\gamma=511$ keV using the geometric area and the photon interaction probability. The Gamma Ray Burst detector consisted of two CsI cylindrical scintillator crystals of 3.8 cm in diameter and 3.2 cm in length~\cite{PVO:1979,PVO:Crider}. Therefore, total geometric area of the detector is nearly $22.7 \text{ cm}^2$. We can obtain the photon interaction probability by using the mass attenuation coefficient for CsI, $(\mu/\rho)|_{511}=0.096 \text{ cm}^2 \text{ g}^{-1}$ and the CsI density $\rho= 4.51  \text{ g}\text{ cm}^{-3}$~\cite{NIST:photon}, 
\begin{equation}
    \text{P}_\text{int}^{511}= 1-\text{Exp}(-4.51\times0.096\times3.2)\simeq 0.75
\end{equation}
Therefore, the effective area is given by 
\begin{equation}
    A_\text{eff}=A_\text{geom}\times\text{P}_\text{int}^{511}=22.7\times0.75  \text{ cm}^2\simeq 17 \text{ cm}^2.
\end{equation}
\begin{center}
\rule{0.3\linewidth}{0.4pt}
\end{center}

\end{document}